\documentclass[conference]{IEEEtran}
\usepackage{balance}
\IEEEoverridecommandlockouts
\usepackage{amsmath,amsfonts}
\usepackage{algorithmic}
\usepackage{algorithm}
\usepackage{array}
\usepackage[caption=false,font=normalsize,labelfont=sf,textfont=sf]{subfig}
\usepackage{textcomp}
\usepackage{stfloats}
\usepackage{url}
\usepackage{verbatim}
\usepackage{graphicx}
\usepackage{cite}
\usepackage{orcidlink}
\usepackage{cancel}
\usepackage{multirow}
\usepackage{adjustbox}
\usepackage{enumitem}
\usepackage{nomencl}
\usepackage{ifthen}
\usepackage{etoolbox}
\usepackage[sort&compress, numbers, square]{natbib}
\usepackage{float}  % Required for \floatname
\floatname{algorithm}{Program}
\allowdisplaybreaks

\def\BibTeX{{\rm B\kern-.05em{\sc i\kern-.025em b}\kern-.08em
    T\kern-.1667em\lower.7ex\hbox{E}\kern-.125emX}}
\usepackage{fancyhdr}
\def\BibTeX{{\rm B\kern-.05em{\sc i\kern-.025em b}\kern-.08em
    T\kern-.1667em\lower.7ex\hbox{E}\kern-.125emX}}
\begin{document}

\title{Stochastic Security Constrained AC Optimal Power Flow Using General Polynomial Chaos Expansion\\
\thanks{This work is supported by the Australian Research in Power Systems Transition (AR-PST) project number (OD-223474) -- This project is led by the Commonwealth Scientific and Industrial Research Organisation (CSIRO).}
\thanks{\noindent Author email: ghulam.mohyuddin@csiro.au}
}
 % , yunqi.wang@csiro.au, rahmat.heidarihaei@powerlink.com.au
\author{\IEEEauthorblockN{
Ghulam Mohy-ud-din\textsuperscript{1,\orcidlink{0000-0003-2738-9338}},~
Yunqi Wang\textsuperscript{1,\orcidlink{0000-0003-1013-5497}},~
Rahmat Heidari\textsuperscript{2,\orcidlink{0000-0003-4835-2614}},~
and Frederik Geth\textsuperscript{3,\orcidlink{0000-0002-9345-2959}}
}
\IEEEauthorblockA{
		$^1${Energy, Commonwealth Scientific and Industrial Research Organisation, Newcastle, NSW, Australia }\\
        $^2${Powerlink, Brisbane, QLD, Australia}\\
        $^3${The University of Queensland, Brisbane, QLD, Australia}\\
}
}

% \IEEEoverridecommandlockouts
% \IEEEpubid{\makebox[\columnwidth]{XXX-X-XXXX-XXXX-X/XX/\$XX.00~\copyright20XX IEEE \hfill} \hspace{\columnsep}\makebox[\columnwidth]{ }}
\maketitle
\thispagestyle{fancy}

% The paper headers
% \markboth{Journal of \LaTeX\ Class Files,~Vol.~14, No.~8, August~2021}%
% {Shell \MakeLowercase{\textit{et al.}}: A Sample Article Using IEEEtran.cls for IEEE Journals}

% \IEEEpubid{0000--0000/00\$00.00~\copyright~2021 IEEE}
% Remember, if you use this you must call \IEEEpubidadjcol in the second
% column for its text to clear the IEEEpubid mark.

\maketitle

\begin{abstract}

Addressing the uncertainty introduced by increasing renewable integration is crucial for secure power system operation, yet capturing it while preserving the full nonlinear physics of the grid remains a significant challenge. This paper presents a stochastic security-constrained optimal power flow model with chance constraints supporting nonlinear AC power flow equations and non-Gaussian uncertainties. We use general polynomial chaos expansion to model arbitrary uncertainties of finite variance, enabling accurate moment computations and robust prediction of system states across diverse operating scenarios. The chance constraints probabilistically limit inequality violations, providing a more flexible representation of controllable variables and the consequent power system operation. Case studies validate the proposed model’s effectiveness in satisfying operational constraints and capturing uncertainty with high fidelity. Compared to the deterministic formulation, it also uncovers a wider set of unsecure contingencies, highlighting improved uncertainty capture and operational insight. 

\end{abstract}


\section*{Nomenclature}

\begin{IEEEdescription}[\IEEEusemathlabelsep\IEEEsetlabelwidth{$\mathcal{I},\, i,j;\ \mathcal{I}_{\mathrm{ref}}$}]

\item[$\mathcal{I},\, i,j;\ \mathcal{I}_{\mathrm{ref}}$] Bus set and index; reference-bus set.
\item[$\mathcal{L},\, {l}$] AC branch set and index.
\item[$\mathcal{T},\, {lij}$] topological set, branch $l$ connects bus $i\!\to\!j$.
\item[$\mathcal{G},\, g;\ \mathcal{G}_i$] Generator set and index; generator at bus i.
\item[$\mathcal{D},\, d;\ \mathcal{D}_i$] Demand set and index; demand at bus $i$.
\item[$\mathcal{K},\, k$] Contingency set and index.
\item[$\mathcal{S},\, s;\ \mathcal{S}_0$] gPCE set and index; gPCE set without $s\!=\!0$.

\item[$\omega,\, \xi(\omega)$] Stochastic germ and mapped standard RV.
\item[$\psi_s(\xi),\, \alpha_s$] Basis polynomial and gPCE coefficient.
\item[$\langle\cdot,\cdot\rangle,\, \Gamma_s,\, \Delta_{rs}$] Inner product, normalize, Kronecker delta.
\item[$\mathsf{M}_{s_1,s_2,s}$] gPCE multiplication tensor.

\item[$\epsilon,\, \lambda(\epsilon)$] Risk level; reformulation factor.
\item[$\sigma$] Standard deviation (uncertain inputs).

\item[$g_l,\, b_l$] Series conductance/susceptance of branch $l$.
\item[$g^{\mathrm{sh}}_i,\, b^{\mathrm{sh}}_i$] Shunt conductance/susceptance at bus $i$.
\item[$\tilde b^{\mathrm{sh}}_i$] Effective shunt susceptance.
% \item[$b^{\mathrm{sh,disp}}_i$] Dispatchable shunt susceptance.
\item[$t_m$] Tap magnitude.

\item[$V^{\mathrm{re}}_{i,s},\, V^{\mathrm{im}}_{i,s}$] Bus-$i$ voltage components.
\item[$W_{i,s}$] Lifted voltage magnitude squared.
\item[$P,\, Q$] Active and reactive power.
\item[$J_{l_{ij},s}$] Lifted current magnitude squared.
\item[$V^{\mathrm{re}}_{lij,s},\, V^{\mathrm{im}}_{l_{ij},s}$] Branch voltage drop components.

\item[$(\cdot)_{s,k}$] Quantity in scenario $s$ under contingency $k$.
\item[$W^+_{i,s,k},\, W^-_{i,s,k}$] PV$\leftrightarrow$PQ bus smoothing auxiliaries.
\item[$\alpha_g$] Active-power participation factor.
\item[$\epsilon$] Smoothing constant (PV/PQ bus switching).

\item[$f(\cdot),\, \psi$] Generation cost; slack penalty weight.
\item[$\mathsf{V,P,Q}$] Sans-serif letters denote uncertain quantities.
\item[$\varsigma^P_g,\, \varsigma^Q_g,\, \varsigma_l$] Slack variables.

% Bounds (unified)
\item[$\underline{(\bullet)},\,\overline{(\bullet)}$] Lower/upper bounds of variables.
\end{IEEEdescription}

\section{Introduction}
% \subsection{Background}
\IEEEPARstart{S}{ecurity}-constrained optimal power flow (SCOPF) is central to reliable and economic system operation, enforcing feasibility in both base and post-contingency states under the $N$–1 criterion \cite{AC_scopf_def}. Addressing the uncertainties involved  in this problem is critical for maintaining the secure operation of the power grid \cite{stochastic_scopf}. This problem inherits two major sources of uncertainties: continuous and discrete. Continuous uncertainty arises from the variable nature of renewable energy sources (RES), while discrete uncertainty stems from contingencies such as outages, making the simultaneous treatment of both within the SCOPF problem a significant modeling challenge. 

% \subsection{Uncertainty Modeling in AC SCOPF problem}

One fundamental challenge in SCOPF arises from the nonlinear and non-convex nature of the AC power flow (AC-PF) equations, which describe the physical relationships among voltages, power injections, and flows within the network \cite{SCOPF_review}. Solving these equations within the SCOPF framework is computationally demanding, particularly for large-scale systems \cite{SCOPF_eff}, which is the second critical challenge for SCOPF.  To improve tractability while managing accuracy, several approximation families are used. The DC linearization remains attractive for its algebraic simplicity and speed, but it neglects voltage and reactive effects \cite{DC}. Convex relaxations provide tighter surrogates such as, SOCP-based SCOPF and quadratic convex (QC) formulations at the cost of larger problem sizes \cite{SOC_relax}, \cite{SOC_relax2}, \cite{QC_relax}. In parallel, {hybrid} representations selectively retain critical AC nonlinearities while simplifying others \cite{ACDC}, and learning-based solvers increasingly assist optimization or act as proxies to expensive subroutines \cite{ML}.

While deterministic SCOPF cast as a two-stage problem separating base and contingency states has been the operational workhorse, its reliance on known demand values limits robustness against variability from RES and load. Stochastic SCOPF addresses this by propagating input uncertainty through the constraints and objective \cite{SCOPF_review3}. Within this family, probabilistic or scenario-based SCOPF accounts for distributions of injections and post-contingency states \cite{prob_SCOPF}, whereas chance-constrained SCOPF enforces violations to be rare with an explicit risk level, often yielding better cost–reliability trade-offs than fully robust designs \cite{chance_SCOPF1}. Robust variants (e.g., CVaR-based) safeguard against extremes but can be overly conservative and expensive \cite{CVaR_robust_SCOCP}.

% Hybrid AC–DC transmission adds another layer of difficulty. HVDC and VSC interfaces enhance long-distance transfers and flexibility, yet their models, DC-side power-flow constraints, and AC–DC couplings enlarge the dimensionality of SCOPF and complicate corrective actions under contingencies \cite{ACDC}. Hence, scalable uncertainty quantification that preserves AC–DC fidelity is essential.

To this end, polynomial chaos expansion (PCE) has emerged as an efficient approach to uncertainty propagation by expanding stochastic inputs on an orthogonal basis and transforming stochastic constraints into deterministic equivalents. It has been applied to optimal power flow problem to reduce sampling cost while maintaining accuracy \cite{PCE}. Generalized PCE (gPCE) further accommodates non-Gaussian inputs and correlated disturbances, improving fidelity and scope \cite{gPCE}, \cite{non_guassian}. Recent work integrates gPCE with chance-constraints to deliver probabilistic guarantees at significantly lower computational cost than the Monte Carlo sampling \cite{gPCE_C_SCOPF}. Furthermore, the trade-off between operational costs and balancing risk under different confidence levels has been efficiently determined using risk-neutral and risk-averse optimal power flow formulations based on non-Gaussian gPCE \cite{risk_gpce}.

% Despite these advancements, most existing studies focus on AC systems and fail to address the unique challenges posed by hybrid AC-DC networks. Further, the introduction of advanced transmission networks (i.e., hybrid AC-DC systems) has further increased the complexity of the SCOPF problem, these networks integrate high-voltage direct current (HVDC) transmission, offer significant advantages for modern power systems, such as enabling long-distance power transfer, enhancing grid stability, and facilitating the large-scale integration of RES \cite{ACDC}. However, the inclusion of DC networks introduces additional challenges, such as voltage source converter (VSC) modeling, DC power flow constraints, and the intricate coupling between AC and DC systems. As these factors significantly increase the dimensionality and computational complexity of SCOPF, the application of stochastic SCOPF to hybrid AC-DC systems remains an open area of research.

In summary, addressing the uncertainty and formulation accuracy challenges in stochastic SCOPF problem is critical for enhancing system security and operational efficiency under renewable energy uncertainty. While existing gPCE-based studies focus exclusively on continuous uncertainty focused on RES, this work, to the best of the authors’ knowledge, is the first to incorporate discrete uncertainties such as $N$–1 contingencies thus, simultaneously addressing the variability of RES and preventively securing outage events. The main contributions of this work are as follows:

\begin{itemize}
    \item Development of a chance-constrained gPCE-based stochastic SCOPF (gPCE-CC-SCOPF) framework, which accounts for uncertainties in RESs and load demands while maintaining system security under $N$-1 contingencies.
   % \item Integration of chance constraints into the SCOPF formulation to provide probabilistic guarantees for operational feasibility. The proposed approach ensures a balance between system reliability and economic efficiency.
   % \item Efficient uncertainty quantification using gPCE to replace computationally expensive MCS. By leveraging gPCE, the proposed framework transforms stochastic constraints into deterministic equivalents, enabling scalable and accurate solutions for large-scale power systems.
   \item Numerical validation on different test systems such as IEEE \texttt{case5}, \texttt{case14}, \texttt{case30} and \texttt{case57} demonstrating the effectiveness of the proposed framework.  
\end{itemize}

The remainder of the paper is organized as follows: Section \ref{gPC_SCOPF} introduces the general polynomial chaos expansion for non-convex stochastic SCOPF, Section \ref{SCOPF} presents the proposed framework, 
% , then Section \ref{Solution} provides the solution methodology, 
Section \ref{Result} provides numerical illustrations; and Section \ref{Conclusion} concludes the paper.

% \item Numerical validation on hybrid test systems to demonstrate the effectiveness of the proposed framework. Comparative studies are conducted against conventional deterministic SCOPF and MCS-based approaches to highlight improvements in computational efficiency, solution accuracy, and system reliability.   

\section{{\normalfont{g}}PCE for Non-convex Stochastic SCOPF} \label{gPC_SCOPF}

\subsection{gPCE Foundations }

In stochastic optimization, uncertain quantities are functions $\mathsf{x}(\omega)$ of a stochastic germ $\omega\in\Omega\subset\mathbb{R}^m$ on $(\Omega,\mathcal{F},\mathbb{P})$. gPCE approximates $\mathsf{x}$ by a truncated orthogonal series of degree $d$:
\begin{align}
\mathsf{x}(\omega)\;\approx\;\hat{\mathsf{x}}(\omega)
=\sum_{s\in\mathcal{S}}\alpha_s\,\psi_s\!\big(\xi(\omega)\big),
\end{align}
where $\xi(\omega)$ is a standardized random vector, $\{\psi_s\}_{s\in\mathcal{S}}$ are polynomials orthogonal with respect to the law of $\xi$, and $\{\alpha_s\}$ are the gPCE coefficients. (The full expansion is infinite for most distributions but not all such as Gaussian; truncation to degree $d$ yields a finite index set $\mathcal{S}$ whose cardinality depends on $m$ and $d$.)

The coefficients follow from projection,
\begin{align}
\alpha_s=\frac{\langle \mathsf{x},\psi_s\rangle}{\langle \psi_s,\psi_s\rangle},
\end{align}
with inner product (expectation) and orthogonality
\begin{align}
\langle \psi_r,\psi_s\rangle
=\mathbb{E}\big[\psi_r(\xi)\psi_s(\xi)\big]
=\Gamma_s\,\Delta_{rs},\qquad \Gamma_s>0 .
\end{align}
Hence,
\begin{align}
\mathbb{E}\big[\hat{\mathsf{x}}\big]=\alpha_0,\qquad
\mathbb{V}\big[\hat{\mathsf{x}}\big]=\sum_{s\in\mathcal{S}\setminus\{0\}}\Gamma_s\,\alpha_s^2 .
\end{align}
Orthogonality enables efficient propagation through precomputed multiplication tensors $\mathsf{M}_{s_1,s_2,s}$  and supports deterministic reformulations such as moment-based chance constraints.

\subsection{ Uncertainty Propagation and Chance Constraints}

% The uncertainty propagation from reference load and generation to nodal voltages and current involves primarily two operations: summation and multiplication. 
In stochastic SCOPF, the propagation of uncertainty plays a critical role in evaluating how input stochastic variables, such as loads or RES, influence system states like voltages, currents, and power flows. These stochastic variables, inherently complex and non-convex, pose significant challenges for optimization due to their irregular distributions and the non-linear nature of the underlying constraints. The gPCE framework offers an effective approach to address these challenges, providing a structured representation for uncertainty propagation while enabling the reformulation of stochastic constraints into deterministic forms.

\begin{figure}[!t]
    \centering
\includegraphics[width=0.9\linewidth]{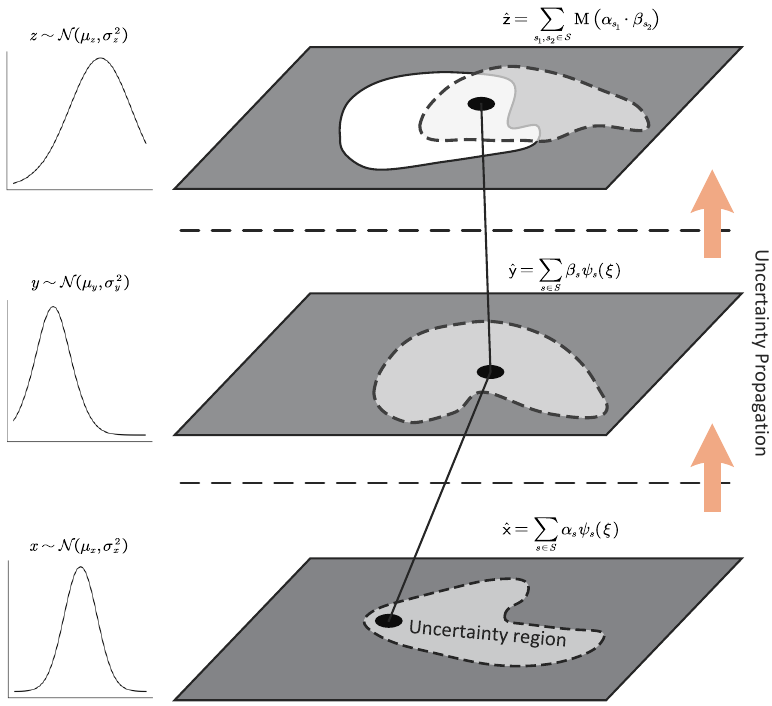}
    \caption{Conceptual uncertainty propagation in stochastic SCOPF. White: deterministic feasible set; black: operating point; gray: variability induced by uncertain inputs. The sketch is qualitative; nonconvex AC constraints can yield irregular feasible regions and paths.}
    \label{fig1}
\end{figure}

Fig. \ref{fig1} illustrates the conceptual process of uncertainty propagation through gPCE. On the right side, the bottom variable $(x)$ represents the input stochastic variable, with its uncertainty visualized as the gray region. As uncertainty propagates to the intermediate variable $(y)$, the gPCE framework approximates $(y)$ in terms of polynomial expansions based on 
$x$, providing a structured representation of uncertainty. Finally, the top variable
$z$ represents the output stochastic variable after further propagation, where the gray region reflects the refined uncertainty range, and the white region highlights a deterministic feasible region. Besides, the left curves show the evolving distributions of 
$x$, $y$, and $z$ during the propagation process. Please be noted the exact distributions may not always be explicitly known, they conceptually demonstrate how uncertainty evolves and becomes structured as it propagates through the gPCE framework.

To mathematically formalize this process, gPCE propagates uncertainty through two primary operations: summation and multiplication, leveraging the Galerkin projection to map stochastic variables onto an orthogonal polynomial basis. For the approximated stochastic variables 
$\hat{\mathsf{x}}$, $\hat{\mathsf{y}}$, and $\hat{\mathsf{z}}$ sharing the same orthogonal basis, these operations are defined as follows:
\begin{align}
    \hat{\mathsf{z}} = \hat{\mathsf{x}} + \hat{\mathsf{y}} \rightarrow \gamma_s = \alpha_s + \beta_s, \forall s \in \mathcal{S}, \\
    \hat{\mathsf{z}} = \hat{\mathsf{x}} \cdot \hat{\mathsf{y}} \rightarrow \gamma_s = \sum\limits_{s_1, s_2 \in \mathcal{S}} \mathsf{M}(\alpha_{s_1} \cdot \beta_{s_2}), \forall s \in \mathcal{S},
\end{align}
where, the pre-computed multiplication tensor $\mathsf{M}$ is, 
\begin{align}
    \mathsf{M} = \frac{\langle \psi_{s_1}, \psi_{s_2}, \psi_{s} \rangle} {\Gamma_s}.
\end{align}

% These operations utilize the pre-computed properties of orthogonal polynomials to simplify the computation of uncertainty propagation, reducing the complexity of stochastic variables into a tractable form through gPCE coefficients.

In addition to propagating uncertainties, gPCE facilitates the reformulation of stochastic constraints into deterministic forms, allowing the optimization problem to remain computationally manageable while preserving probabilistic characteristics. Chance constraints are employed to ensure that the probability of exceeding deterministic bounds $\overline{\mathsf{x}}$, $\underline{\mathsf{x}}$ remains below a specified tolerance level $\epsilon$. These constraints are expressed as:
\begin{align}
\begin{aligned}
    \mathbb{P} (\mathsf{x} \geq \underline{\mathsf{x}}) \geq (1-\epsilon),\\
    \mathbb{P} (\mathsf{x} \leq \overline{\mathsf{x}}) \geq (1-\epsilon).
\end{aligned}
\end{align}
and reformulated using the expected value and variance as:
\begin{align}
    \underline{\mathsf{x}} \leq \mathbb{E} (\mathsf{x}) \pm \lambda(\epsilon) \sqrt{\mathbb{V}(\mathsf{x})} \leq \overline{\mathsf{x}}.
\end{align}

Here, for the choice of $\epsilon$, and the reformulation factor $\lambda(\epsilon)$ for all limits such as for each node $i$, $\lambda_i(\epsilon^{\mathsf{V}})$ and for each branch $l$  connecting bus $i$ and $j$, there is a topological triple $lij$, and a $\lambda_{lij}(\epsilon^{\mathsf{I}})$, empirical tuning technique is used \cite{gPCE_C_SCOPF}. Further, in this formulation, the fourth-order tensor product is reduced to second-order by using a few auxiliary variables.

% \begin{align}
% \mathsf{W}_{i,s} = (\mathsf{V}_{i,s}^{\mathsf{re}}^2 + \mathsf{V}_{i,s}^{\mathsf{im}}^2),
% \end{align}

\section{Proposed Stochastic SCOPF Framework}\label{SCOPF}

% \subsection{Framework Overview}
The proposed stochastic gPCE-CC-SCOPF is formulated in the power-voltage rectangular coordinate system, as a two-stage mathematical programming problem. The first stage represents the gPCE-based base case formulation with chance constraints and the second stage represents the deterministic contingency-cases linked to each base case scenario.

\subsection{Objective Function} \label{subsec:obj}
The objective function minimizes the expected active generation cost and the penalized chance constraint violations, 
\begin{multline}\label{eq:obj}     
    \min~~\sum_{g \in \mathcal G} \mathbb{E} (f(\overline{\mathsf{P_g}})) + \psi~\varsigma^{P}_{g}~+~\psi~\varsigma^{Q}_{g}~+~\sum_{l \in \mathcal{L}} \psi~\varsigma_{l}.
\end{multline}
\subsection{Base Case Constraints} \label{subsec:basecase}
The base case provides a feasible set of constraints for the non-contingent gPCE based chance constrained power flow formulation for each scenario $s$, where the chance constraints limit constraint violation probabilities.  
\subsubsection{Constraints for reference bus $i, \forall i \in \mathcal{I}^{\mathsf{ref}}$} 
\begin{align}
    \mathsf{V}_{i,s}^{\mathsf{re}} &= 0, \forall s \in \mathcal{S}_0, \label{eq:b_ref} \\
    \mathsf{V}_{i,s}^{\mathsf{im}} &= 0, \forall s \in \mathcal{S}. \label{eq:bs_ref}
\end{align}
The constraints \eqref{eq:b_ref} and \eqref{eq:bs_ref} sets the reference bus real and imaginary voltage.
\subsubsection{Constraints for generator $g, \forall g \in \mathcal{G}$}  
\begin{align} 
   \underline{P_g} \leq \mathbb{E}(\overline{\mathsf{P}_g}) \pm \lambda(\epsilon^P) \sqrt{\mathbb{V}(\overline{\mathsf{P}_g})} \leq \overline{P_g} + \varsigma^{P}_{g}, \label{eq:pbounds_cc} \\
   \underline{Q_g} \leq \mathbb{E}(\overline{\mathsf{Q}_g}) \pm \lambda(\epsilon^Q) \sqrt{\mathbb{V}(\overline{\mathsf{Q}_g})} \leq \overline{Q_g}+ \varsigma^{Q}_{g}. \label{eq:qbounds_cc}
\end{align}
The chance constraints \eqref{eq:pbounds_cc} and \eqref{eq:qbounds_cc} ensure that the probability of stochastic active and reactive power generation violating the generation limits is less than the tolerance levels while allowing for violations using slack variables. 
\subsubsection{Constraints for bus $i$, $\forall i \in \mathcal{I}$}
\begin{gather}
    \mathsf{W}_{i,s} = \sum\limits_{s_1, s_2 \in \mathcal{S}} \mathsf{M}(\mathsf{V}_{i,s_1}^{\mathsf{re}} \mathsf{V}_{i,s_2}^{\mathsf{re}} +  \mathsf{V}_{i,s_1}^{\mathsf{im}}  \mathsf{V}_{i,s_2}^{\mathsf{im}}), \forall s \in \mathcal{S}, \label{eq:lifted_v} \\
    \sum\limits_{g \in \mathcal{G}_i} \mathsf{P}_{g,s} - \!\!\!\sum\limits_{lij \in \mathcal{L}_i} \mathsf{P}_{lij,s} - g_{i}^{\mathsf{sh}} \mathsf{W}_{i,s} \! = \sum\limits_{d \in \mathcal{D}_i} P_{d,s}, \forall s \in \mathcal{S}, \label{eq:p_balance} \\
    \sum\limits_{g \in \mathcal{G}_i} \mathsf{Q}_{g,s} \!-\! \sum\limits_{lij \in \mathcal{L}_i} \mathsf{Q}_{lij,s} + {{b}_{i}^{\mathsf{sh}}} \mathsf{W}_{i,s}
   \!=\! \sum\limits_{d \in \mathcal{D}_i} Q_{d,s}, \forall s \in \mathcal{S}, \label{eq:q_balance} \\  
    % \underline{b_{i}^{\mathsf{sh,disp}}} \leq b_{i,s}^{\mathsf{sh,disp}} \leq \overline{b_{i}^{\mathsf{sh,disp}}}, \forall s \in \mathcal{S}, \label{eq:shunt_dispatch_limits} \\
    \underline{V_i}^2 \leq \mathbb{E}(\overline{\mathsf{W_i}}) \pm \lambda(\epsilon^V) \sqrt{\mathbb{V}(\overline{\mathsf{W_i}})} \leq \overline{V_i}^2. \label{eq:vbounds_cc} 
\end{gather}
The constraints \eqref{eq:lifted_v} links the lifted to the real voltage variables, \eqref{eq:p_balance} and \eqref{eq:q_balance} implements the active and reactive power balance, 
% \eqref{eq:shunt_dispatch_limits} provides bounds to the dispatchable shunts,
and \eqref{eq:vbounds_cc} ensures that the probability of lifted voltage limit violations remain below given tolerance level. 

% Finally, in \eqref{eq:q_balance}, ${\tilde{b}_{i}^{\mathsf{sh}}} = {{b}_{i}^{\mathsf{sh}}} + {{b}_{i}^{\mathsf{sh,disp}}}$.

\subsubsection{Constraints for ac branch $l, \forall lij \in \mathcal{T}$}
\begin{gather}
    \mathsf{P}_{lij,s} = \sum\limits_{s_1, s_2 \in \mathcal{S}} \mathsf{M} \bigg( \frac{(g_l + g_{lij})}{(t^{\text{m}}_l)^2} (\mathsf{V}_{i,s_1}^{\mathsf{re}} \mathsf{V}_{i,s_2}^{\mathsf{re}} +  \mathsf{V}_{i,s_1}^{\mathsf{im}}  \mathsf{V}_{i,s_2}^{\mathsf{im}}) \nonumber \\ 
    + \frac{(b_l t^{\text{im}}_l - g_l t^{\text{re}}_l)}{(t^{\text{m}}_l)^2} (\mathsf{V}_{i,s_1}^{\mathsf{re}} \mathsf{V}_{j,s_2}^{\mathsf{re}} +  \mathsf{V}_{i,s_1}^{\mathsf{im}}  \mathsf{V}_{j,s_2}^{\mathsf{im}}) \nonumber  \\ 
    - \frac{(b_l t^{\text{re}}_l + g_l t^{\text{im}}_l )}{(t^{\text{m}}_l)^2} (\mathsf{V}_{i,s_1}^{\mathsf{re}}  \mathsf{V}_{j,s_2}^{\mathsf{im}} - \mathsf{V}_{i,s_1}^{\mathsf{im}} \mathsf{V}_{j,s_2}^{\mathsf{re}}) \bigg), \forall s \in \mathcal{S}  \label{eq:pij_flow} \\
    \mathsf{Q}_{lij,s} = \sum\limits_{s_1, s_2 \in \mathcal{S}} \mathsf{M} \bigg( \frac{-(b_l + b_{lij})}{(t^{\text{m}}_l)^2} (\mathsf{V}_{i,s_1}^{\mathsf{re}} \mathsf{V}_{i,s_2}^{\mathsf{re}} +  \mathsf{V}_{i,s_1}^{\mathsf{im}}  \mathsf{V}_{i,s_2}^{\mathsf{im}}) \nonumber \\
    - \frac{(-b_l t^{\text{re}}_l - g_l t^{\text{im}}_l )}{(t^{\text{m}}_l)^2} (\mathsf{V}_{i,s_1}^{\mathsf{re}} \mathsf{V}_{j,s_2}^{\mathsf{re}} +  \mathsf{V}_{i,s_1}^{\mathsf{im}}  \mathsf{V}_{j,s_2}^{\mathsf{im}}) \nonumber  \\
    + \frac{(b_l t^{\text{im}}_l  - g_l t^{\text{re}}_l)}{(t^{\text{m}}_l)^2} (\mathsf{V}_{i,s_1}^{\mathsf{im}}  \mathsf{V}_{j,s_2}^{\mathsf{re}} - \mathsf{V}_{i,s_1}^{\mathsf{re}} \mathsf{V}_{j,s_2}^{\mathsf{im}}) \bigg), \forall s \in \mathcal{S}  \label{eq:qij_flow} \\ 
    \mathsf{P}_{lji,s} = \sum\limits_{s_1, s_2 \in \mathcal{S}} \mathsf{M} \bigg( \frac{(g_l + g_{lji})}{(t^{\text{m}}_l)^2} (\mathsf{V}_{j,s_1}^{\mathsf{re}} \mathsf{V}_{j,s_2}^{\mathsf{re}} +  \mathsf{V}_{j,s_1}^{\mathsf{im}}  \mathsf{V}_{j,s_2}^{\mathsf{im}}) \nonumber \\
    + \frac{(-g_l t^{\text{re}}_l - b_l t^{\text{im}}_l )}{(t^{\text{m}}_l)^2} (\mathsf{V}_{i,s_1}^{\mathsf{re}} \mathsf{V}_{j,s_2}^{\mathsf{re}} +  \mathsf{V}_{i,s_1}^{\mathsf{im}}  \mathsf{V}_{j,s_2}^{\mathsf{im}}) \nonumber \\
    + \frac{(g_l t^{\text{im}}_l  - b_l t^{\text{re}}_l)}{(t^{\text{m}}_l)^2} (\mathsf{V}_{i,s_1}^{\mathsf{re}}  \mathsf{V}_{j,s_2}^{\mathsf{im}} - \mathsf{V}_{i,s_1}^{\mathsf{im}} \mathsf{V}_{j,s_2}^{\mathsf{re}}) \bigg), \forall s \in \mathcal{S}  \label{eq:pji_flow} \\
    \mathsf{Q}_{lji,s} = \sum\limits_{s_1, s_2 \in \mathcal{S}} \mathsf{M} \bigg( \frac{-(b_l + b_{lji})}{(t^{\text{m}}_l)^2} (\mathsf{V}_{j,s_1}^{\mathsf{re}} \mathsf{V}_{j,s_2}^{\mathsf{re}} +  \mathsf{V}_{j,s_1}^{\mathsf{im}}  \mathsf{V}_{j,s_2}^{\mathsf{im}}) \nonumber \\
    + \frac{(g_l t^{\text{im}}_l  - b_l t^{\text{re}}_l)}{(t^{\text{m}}_l)^2} (\mathsf{V}_{i,s_1}^{\mathsf{re}} \mathsf{V}_{j,s_2}^{\mathsf{re}} +  \mathsf{V}_{i,s_1}^{\mathsf{im}}  \mathsf{V}_{j,s_2}^{\mathsf{im}}) \nonumber \\
    - \frac{(g_l t^{\text{re}}_l+ b_l t^{\text{im}}_l )}{(t^{\text{m}}_l)^2} (\mathsf{V}_{i,s_1}^{\mathsf{im}}  \mathsf{V}_{j,s_2}^{\mathsf{re}} - \mathsf{V}_{i,s_1}^{\mathsf{re}} \mathsf{V}_{j,s_2}^{\mathsf{im}}) \bigg), \forall s \in \mathcal{S}  \label{eq:qji_flow} \\
    \mathsf{J}_{lij,s} = (g_l^2 + b_l^2) \hspace{6cm}   \nonumber \\
    \sum\limits_{s_1, s_2 \in \mathcal{S}} \mathsf{M} (\mathsf{V}_{lij,s_1}^{\mathsf{re}} \mathsf{V}_{lij,s_2}^{\mathsf{re}} + \mathsf{V}_{lij,s_1}^{\mathsf{im}} \mathsf{V}_{lij,s_2}^{\mathsf{im}} ), \forall s \in \mathcal{S}  \label{eq:Jij_flow} \\
    \mathsf{V}_{lij,s}^{\text{re}} = \mathsf{V}_{i,s}^{\text{re}} - \mathsf{V}_{j,s}^{\text{re}}, \forall s \in \mathcal{S} \label{eq:vij} \\
    \mathsf{V}_{lij,s}^{\text{im}} = \mathsf{V}_{i,s}^{\text{im}} - \mathsf{V}_{j,s}^{\text{im}}, \forall s \in \mathcal{S} \label{eq:vji} \\
    \mathbb{E}(\overline{\mathsf{J_{lij}}}) \pm \lambda(\epsilon^{\text{I}}) \sqrt{\mathbb{V}(\overline{\mathsf{J_{lij}}})} \leq \overline{I_l^2} + \varsigma_{l}. \label{eq:Jij_flow_cc}
\end{gather}
The constraints \eqref{eq:pij_flow}, \eqref{eq:qij_flow}, \eqref{eq:pji_flow}, and \eqref{eq:qji_flow} represent the active and reactive power flows, \eqref{eq:Jij_flow} determines the lifted current flow, \eqref{eq:vij} and \eqref{eq:vji} determines the voltage drop of ac branches. Finally, \eqref{eq:Jij_flow_cc} ensures that the probability of lifted current flow limits violation on ac branches remains within the prescribed threshold.
\subsection{Contingency Case Constraints} \label{subsec:contcase}
The contingency case $k$ provides a feasible set of constraints for the contingent deterministic power flow formulation linked to each base case scenario $s$.
\subsubsection{Constraints for reference bus $i, \forall i \in \mathcal{I}^{\mathsf{ref}}$}  
\begin{align}
    \mathsf{V}_{i,s, k}^{\mathsf{im}} = 0, \forall k \in \mathcal{K}. \label{eq:c_ref}
\end{align}
The constraint \eqref{eq:c_ref} sets the reference phasor to 0.
\subsubsection{Constraints for bus $i, \forall i \in \mathcal{I}, \forall k \in \mathcal{K} $ }
\begin{gather}
    {W}_{i,s,k} = ({V}_{i,s,k}^{\mathsf{re}})^2 + ({V}_{i,s,k}^{\mathsf{im}})^2, \label{eq:c_lifted_v} \\
    \sum\limits_{g \in \mathcal{G}_i} {P}_{g,s,k} - \!\!\!\sum\limits_{lij \in \mathcal{L}_i} {P}_{lij,s,k} - g_{i}^{\mathsf{sh}} \mathsf{W}_{i,s} \! = \sum\limits_{d \in \mathcal{D}_i} P_{d,s },  \label{eq:c_p_balance} \\
    \sum\limits_{g \in \mathcal{G}_i} {Q}_{g,s,k} \!-\! \sum\limits_{lij \in \mathcal{L}_i} {Q}_{lij,s,k} + {{b}_{i}^{\mathsf{sh}}} \mathsf{W}_{i,s} = \sum\limits_{d \in \mathcal{D}_i} Q_{d,s}.  \label{eq:c_q_balance} 
\end{gather}
The constraints \eqref{eq:c_lifted_v} links the lifted to the real voltage variables, and \eqref{eq:c_p_balance} and \eqref{eq:c_q_balance} implements the active and reactive power balance.
\subsubsection{Constraints for ac branches, $\forall lij \in \mathcal{T}, \forall k \in \mathcal{K}$}
\begin{gather}
    {P}_{lij,s,k} = \frac{(g_l + g_{lij})}{(t^{\text{m}}_l)^2} ({W}_{i,s,k}) \hspace{4.3cm} \nonumber \\ 
    + \frac{(b_l t^{\text{im}}_l - g_l  t^{\text{re}}_l)}{(t^{\text{m}}_l)^2} ({V}_{i,s,k}^{\mathsf{re}} {V}_{j,s,k}^{\mathsf{re}} +  {V}_{i,s,k}^{\mathsf{im}}  {V}_{j,s,k}^{\mathsf{im}}) \nonumber  \\ 
    - \frac{(b_l  t^{\text{re}}_l + g_l t^{\text{im}}_l)}{(t^{\text{m}}_l)^2} ({V}_{i,s,k}^{\mathsf{re}}  {V}_{j,s,k}^{\mathsf{im}} - {V}_{i,s,k}^{\mathsf{im}} {V}_{j,s,k}^{\mathsf{re}}) ,  \label{eq:c_pij_flow} \\
    {Q}_{lij,s,k} = \frac{-(b_l + b_{lij})}{(t^{\text{m}}_l)^2} ({W}_{i,s,k}) \hspace{4.3cm} \nonumber \\ 
    - \frac{(-b_l  t^{\text{re}}_l - g_l t^{\text{im}}_l)}{(t^{\text{m}}_l)^2} ({V}_{i,s,k}^{\mathsf{re}} {V}_{j,s,k}^{\mathsf{re}} + {V}_{i,s,k}^{\mathsf{im}}  {V}_{j,s,k}^{\mathsf{im}}) \nonumber  \\
    + \frac{(b_l t^{\text{im}}_l - g_l  t^{\text{re}}_l)}{(t^{\text{m}}_l)^2} ({V}_{i,s,k}^{\mathsf{im}}  {V}_{j,s,k}^{\mathsf{re}} - {V}_{i,s,k}^{\mathsf{re}} {V}_{j,s,k}^{\mathsf{im}}) ,  \label{eq:c_qij_flow}\\ 
    {P}_{lji,s,k} = \frac{(g_l + g_{lji})}{(t^{\text{m}}_l)^2} ({W}_{j,s,k}) \hspace{4.3cm} \nonumber \\
    + \frac{(-g_l  t^{\text{re}}_l - b_l t^{\text{im}}_l)}{(t^{\text{m}}_l)^2} ({V}_{i,s,k}^{\mathsf{re}} {V}_{j,s,k}^{\mathsf{re}} +  {V}_{i,s,k}^{\mathsf{im}}  {V}_{j,s,k}^{\mathsf{im}}) \nonumber \\
    + \frac{(g_l t^{\text{im}}_l - b_l  t^{\text{re}}_l)}{(t^{\text{m}}_l)^2} ({V}_{i,s,k}^{\mathsf{re}}  {V}_{j,s,k}^{\mathsf{im}} - {V}_{i,s,k}^{\mathsf{im}} {V}_{j,s,k}^{\mathsf{re}}) ,  \label{eq:c_pji_flow} \\
    {Q}_{lji,s,k} = \frac{-(b_l + b_{lji})}{(t^{\text{m}}_l)^2} ({W}_{j,s,k}) \hspace{4.3cm} \nonumber \\
    + \frac{(g_l t^{\text{im}}_l - b_l  t^{\text{re}}_l)}{(t^{\text{m}}_l)} ({V}_{i,s,k}^{\mathsf{re}} {V}_{j,s,k}^{\mathsf{re}} + {V}_{i,s,k}^{\mathsf{im}}  {V}_{j,s,k}^{\mathsf{im}}) \nonumber \\
    - \frac{(g_l  t^{\text{re}}_l + b_l t^{\text{im}}_l)}{(t^{\text{m}}_l)^2} ({V}_{i,s,k}^{\mathsf{im}}  {V}_{j,s,k}^{\mathsf{re}} - {V}_{i,s,k}^{\mathsf{re}} {V}_{j,s,k}^{\mathsf{im}}) ,  \label{eq:c_qji_flow}  \\ 
    {J}_{lij,s,k} = (g_l^2 + b_l^2) 
     (({V}_{lij,s,k}^{\mathsf{re}})^2 + ({V}_{lij,s,k}^{\mathsf{im}})^2 ), \qquad  \label{eq:c_Jij_flow}  \\
    {V}_{lij,s,k}^{\text{re}} = {V}_{i,s,k}^{\text{re}} - {V}_{j,s,k}^{\text{re}}, \qquad \qquad \quad  \label{eq:c_vij}\\
    {V}_{lij,s,k}^{\text{im}} = {V}_{i,s,k}^{\text{im}} - {V}_{j,s,k}^{\text{im}}. \qquad \qquad \quad  \label{eq:c_vji}
\end{gather}
The constraints \eqref{eq:c_pij_flow}, \eqref{eq:c_qij_flow}, \eqref{eq:c_pji_flow}, and \eqref{eq:c_qji_flow} represent the active and reactive power flows, \eqref{eq:c_Jij_flow} determines the lifted current flow, and \eqref{eq:c_vij} and \eqref{eq:c_vji} determines the voltage drop in ac branches.
\subsubsection{PQ Response Constraints, $\forall k \in \mathcal{K}$}
\begin{gather}
    {P}_{g,s,k} =  \mathsf{P}_{g,s}+\alpha_g \Delta_{s,k}, \forall g \in \mathcal{G}^{\text{resp}},   \label{eq:p_gen_response} \\
    {P}_{g,s,k} =  \mathsf{P}_{g,s}, \forall g \in \mathcal{G}^{\text{non-resp}},  \label{eq:p_gen_no_response} \\
    W_{i,s,k} = \mathsf{W}_{i,s} + \epsilon \ln \bigg(1 + \exp^{\big((W_{i,s,k}^+ - Q_{g,s,k} + \underline{Q_{g}} )/\epsilon\big)} \bigg) \quad \nonumber \\ 
- \epsilon \ln \bigg(1 + \exp^{\big((W_{i,s,k}^- + Q_{g,s,k} - \overline{Q_{g}} )/\epsilon\big)}\bigg), \forall gi \in \mathcal{G} \times \mathcal{I},  \label{eq:q_gen_response}\\
    0 \leq W_{i,s,k}^+ \leq \overline{V_i}^2 - \mathsf{W}_{i,s}, \;\; \forall gi \in \mathcal{G} \times \mathcal{I}, \label{eq:add_smooth1}\\
    0 \leq W_{i,s,k}^- \leq \mathsf{W}_{i,s} - \underline{V_i}^2, \;\; \forall gi \in \mathcal{G} \times \mathcal{I}. \label{eq:add_smooth2}
 \end{gather}
The active power of generators responding and non-responding generators in contingency $k$ is given by \eqref{eq:p_gen_response} and \eqref{eq:p_gen_no_response} respectively. The generators maintain the base case reference voltage by changing their reactive power until they hit operation limits. This is referred to as PV/PQ bus switching control and it is modeled here in \eqref{eq:q_gen_response} using smooth approximation as given in \cite{smooth_approx}. Finally, \eqref{eq:add_smooth1} and \eqref{eq:add_smooth2} are supporting constraints for \eqref{eq:q_gen_response}.

\section{Numerical Illustration} \label{Result}

\subsection{Deterministic Algebraic Optimization Model} \label{Solution}

This proposed gPCE-CC-SCOPF model is solved as a multi-network optimization as shown in program \ref{alg:mn_opt}. The sets of gPCE scenarios ($\mathcal{S} = \{s_1, \ldots, s_{n}\}$) and contingencies are clustered to formulate a multi-network dataset representing the networks such that $s_1, \ldots, s_{n}, k_{s1,1}, \ldots, k_{s1,n}, \dots k_{sn,1}, \ldots, k_{sn,n}$. Here, the mapping of each scenario to their respective contingency sets is provided by $\mathcal{K}_{sn}$ sets. The optimization model is built such that each scenario sets base case decision variables and constraints represented by \eqref{eq:b_ref}-\eqref{eq:Jij_flow_cc} defined in section \ref{subsec:basecase} followed by the decision variables and constraints of each linked contingency case represented by \eqref{eq:c_ref}-\eqref{eq:add_smooth2} defined in section \ref{subsec:contcase} including the PQ response coupling constraints. Then the objective function represented by \eqref{eq:obj} given in section \ref{subsec:obj} is defined.

\begin{algorithm}[!tp]
\caption{gPCE-CC-SCOPF Multi-Network Optimization} \label{alg:mn_opt}
\begin{algorithmic}[1]
\STATE \textbf{Input:} multi-network dataset, \\
\qquad~~$\{s_1, \ldots, s_{n}, k_{s1,1}, \ldots, k_{s1,n}, \dots k_{sn,1}, \ldots, k_{sn,n}\},$ \\
\quad$\mathcal{S} = \{s_1, \ldots, s_{n}\}, \mathcal{K}_{s1} =\{k_{s1,1}, \ldots, k_{s1,n} \}, \dots, \mathcal{K}_{sn}.$ \\
\STATE \textbf{Output:} Stochastic preventive optimal solution

\STATE Initialize objective function $f(\cdot)$

\FOR{each $s \in \mathcal{S}$}
    \STATE Define base case decision variables $x_s$
    \STATE Apply base case constraints for $s$: $C_s(x_s)$
    
    \FOR{each contingency $k \in K_s$}
        \STATE Define contingency case variables $x_{s,k}$
        \STATE Apply contingency case constraints $C_{s,k}(x_{s,k})$
        \STATE Apply PQ response constraints $C_{s,k}(x_s, x_{s,k})$
    \ENDFOR
\ENDFOR

\STATE Solve: $\min f(x_s, x_{s,k})$ subject to all constraints

\end{algorithmic}
\end{algorithm}

\subsection{Implementation and Computational Setup}
This optimization model is built using JuMP in Julia and solved by the Ipopt solver (v1.2.0). 
The code is available open-source on GitHub with the test systems data \footnote{https://github.com/Electa-Git/StochasticPowerModels.jl}. All simulations were performed on  an Apple M4 Pro chip (14‑core CPU, 20‑core GPU), 48 GB unified memory and macOS Sequoia 15.3.2. The nonlinear program \eqref{eq:obj}-\eqref{eq:add_smooth2} was initialized by assigning the zero-order gPCE coefficients to match the outcome of the deterministic SCOPF problem, evaluated using the expected values of the uncertain parameters. All higher-order coefficients were initially set to zero. Additionally, the deterministic SCOPF solution served as a preliminary estimate for the active constraint set in \eqref{eq:obj}-\eqref{eq:add_smooth2}.

\subsection{Case Study Data}
In this section, the application of the proposed gPCE-based chance-constrained SCOPF for various test cases including IEEE \texttt{case5}, \texttt{case14}, \texttt{case30}, and, \texttt{case57} is shown. In this study we considered the standard deviation $\sigma$ and risk level $\epsilon$ equal to 0.10. This section provides a comparison with deterministic counterpart and discusses the limitations and applications of this tool. 

\subsection{Proof-of-concept Study}

\begin{table*}[h!]
\caption{Comparison of Deterministic and Proposed \normalfont{g}PCE-CC-SCOPF Models}
\centering
\begin{adjustbox}{max width=\textwidth}
\begin{tabular}{ l c c c c c c c c c c c}
\hline 
\multirow{3}{*}{\bf Case}   & \multicolumn{2}{c}{ \bf Contingencies} & \multicolumn{3}{c}{\bf Deterministic SCOPF Model}  & \multicolumn{3}{c}{\bf gPCE-CC-SCOPF Model (deg=1)} & \multicolumn{3}{c}{\bf gPCE-CC-SCOPF Model (deg=2)} \\ \cline{2-3} \cline{4-6} \cline{7-9} \cline{10-12}  
                            & \bf Generator & \bf Branch & \bf Objective & \bf Unsecure   & \bf Time & \bf Objective & \bf Unsecure & \bf Time & \bf Objective & \bf Unsecure & \bf Time \\  
                            & (gx) & (bx)  &  (\$/h)  &  (-)  &  (s)  & (\$/h)   & (-) & (s) & (\$/h)   & (-) & (s) \\ \hline
\texttt{case5}     & 4  & 7     & 922947.89 & -           & 0.00693   & 926655.74 & -  &  0.060973 & 926655.74 & -  &  0.09716 \\ 
\texttt{case14}    & 5  & 19    & 2178.08 & b1          & 0.00964   & 2195.69 & b1  & 0.133604 & 2195.70 & b1  & 5.20927 \\ 
\texttt{case30}    & 6  & 38    & 626.25 & g3, g6, b30 & 0.57122  & 631.09  & g3, g6, b30 & 1.1784 & 633.89  & g3, g6, b30, b32 & 71.1784 \\ 
\texttt{case57}    & 7  & 79    & 37589.33   & b7, b22   & 0.61672  & 37629.14  & b7, b22 & 1.81088 & 37630.52  & g5, b7, b8, b22  & 96.7562 \\ 
\hline
\end{tabular}
\end{adjustbox}
\label{tab:comparison}
\end{table*}

Table \ref{tab:comparison} highlights the comparative performance between the traditional deterministic SCOPF and the proposed gPCE-CC-SCOPF models using multiple test cases. The proposed formulation with degree ($d$) 1, and 2 introduce modest increases in cost and computation time, while demonstrating improved robustness by enhancing the contingency awareness and operation security under uncertainty. Particularly in \texttt{case30} and \texttt{case57}, the proposed method with degree 2 identified a broader set of unsecure contingencies by accurately capturing the load uncertainty, which remained undetected under the deterministic formulation.

\begin{figure}[t]
    \centering
    \includegraphics[width=0.95\columnwidth]{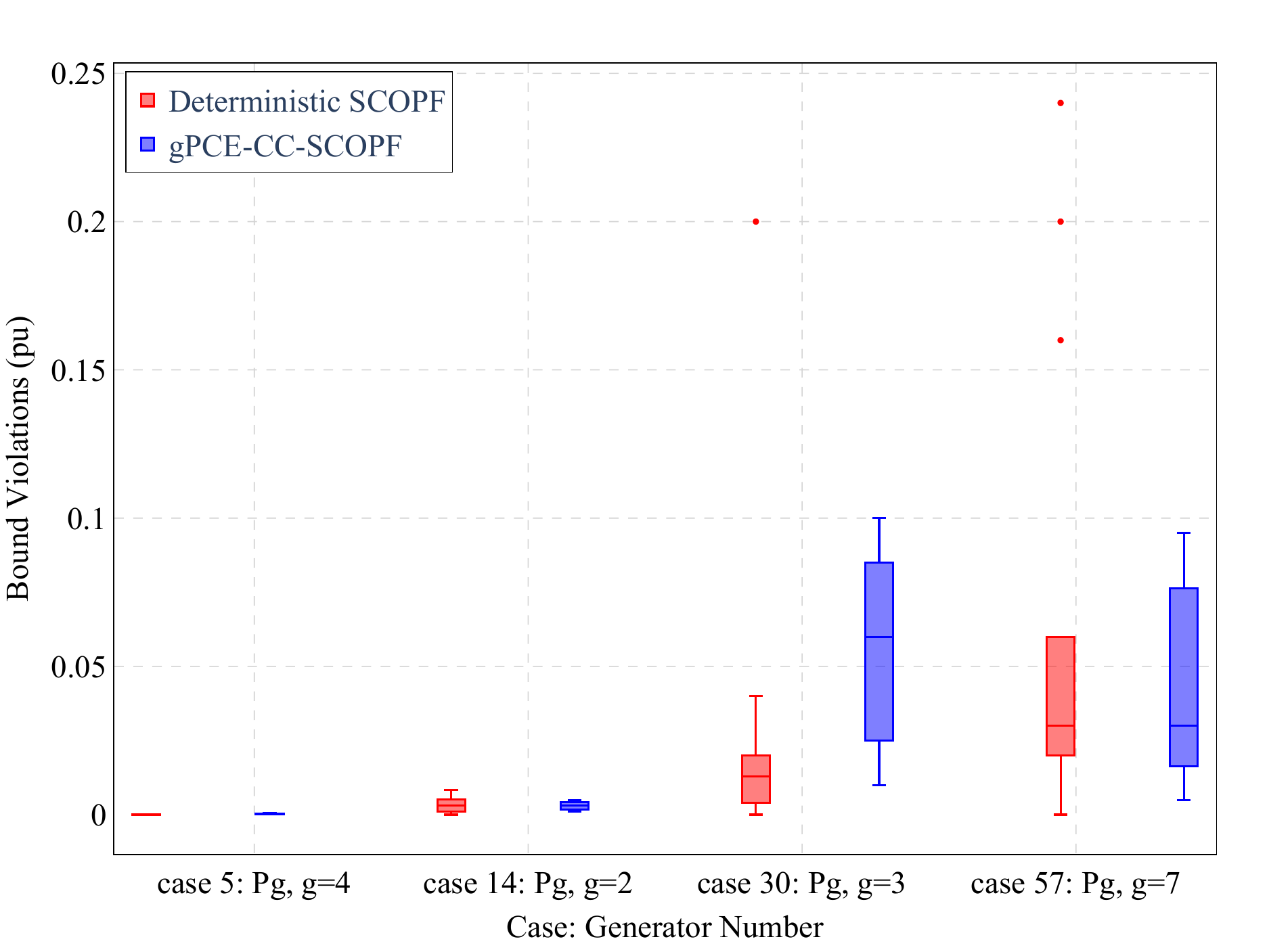} 
    \caption{Generator active power bound violation profile under contingency set.}
    \label{fig:violation_Pg}
\end{figure}

Next, the in-sample capability of the proposed method to suppress constraint violations to within acceptable bounds is examined and compared with the deterministic counterpart. Fig. \ref{fig:violation_Pg} illustrates a comparative assessment of generator active power bound violations under the deterministic and the proposed formulation across four test systems by choosing different generators. Across all systems, the proposed formulation consistently yields lower or comparable violation distributions, especially pronounced in \texttt{case30} and \texttt{case57}, where maximum violation levels in some cases for the deterministic model reach or exceed 0.2 pu (indicated by the outliers). In contrast, the proposed formulation restricts violations well below 0.1 pu, suggesting improved probabilistic constraint handling. Furthermore, the reduced inter-quartile ranges and absence of strong outliers in the gPCE traces indicate enhanced robustness and tighter control across uncertain scenarios.

\begin{figure}[htbp]
    {\centering   
    \includegraphics[width=0.95\columnwidth]{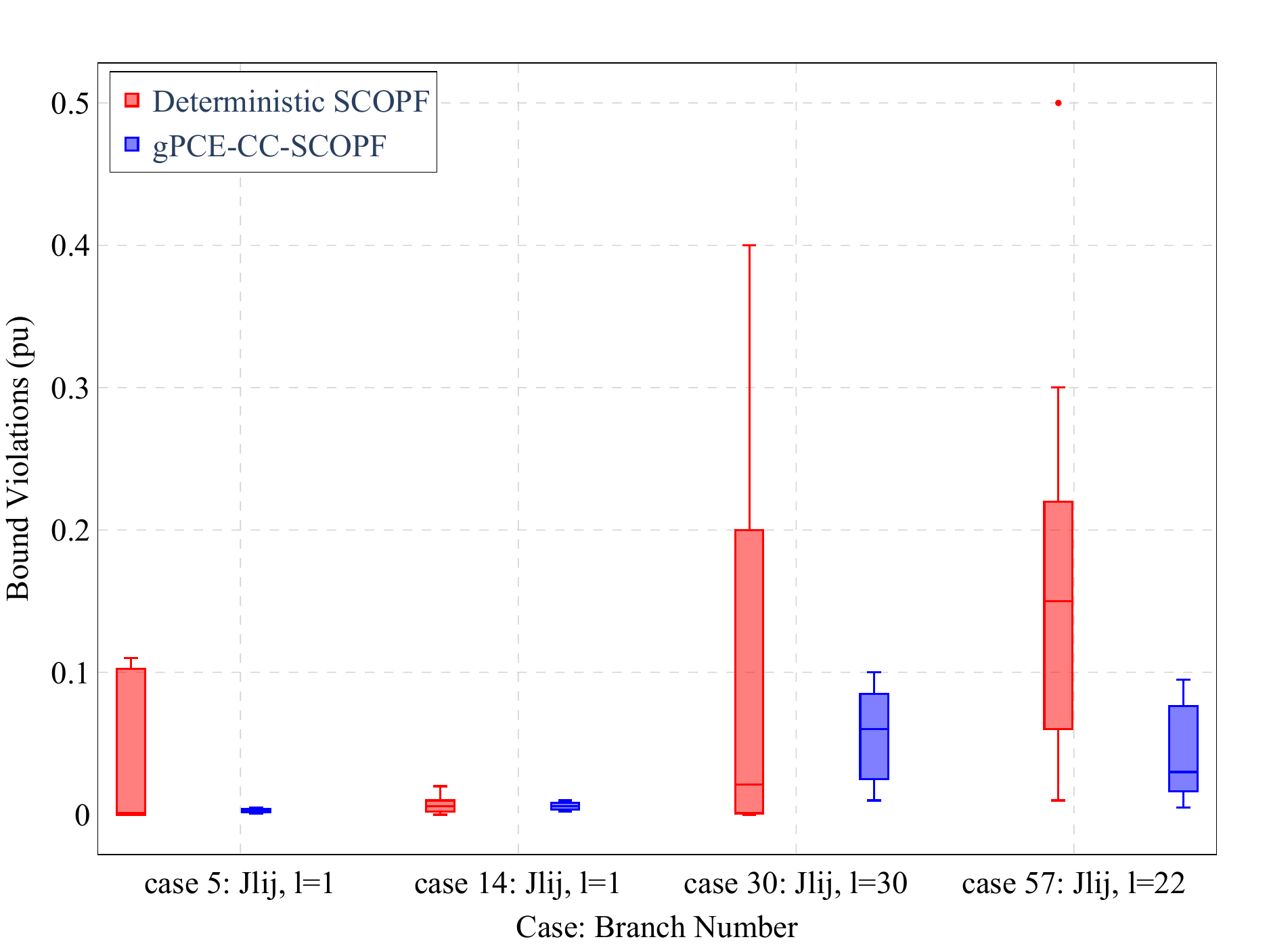} }
    \label{fig:violation_Jlij}
    \caption{Branch current limit violation profile under contingency set.}
\end{figure}

Similarly, Fig. \ref{fig:violation_Jlij} presents the lifted branch current bound violations for deterministic and proposed formulations. The box plots demonstrate that the deterministic model tends to produce significantly higher and more variable violations, particularly in \texttt{case30} and \texttt{case57}, the values exceed 0.4 pu. In contrast, the proposed formulation shows a marked reduction in both violation magnitude and spread, with all values confined well below the critical 0.1 pu threshold indicating enhanced constraint satisfaction and probabilistic robustness.

\section{Conclusion} \label{Conclusion}
This paper presents a tractable stochastic chance-constrained AC-SCOPF model using gPCE, enabling direct handling of the full nonlinear AC power flow equations without relying on sampling, linearizations, or relaxations. Evaluated across test systems, the method consistently restricts constraint violations and enhances solution system security under uncertainty by picking up a broader set of unsecure contingencies, which remained undetected under the deterministic formulation. The results indicate that the proposed formulation provides sufficiently accurate solutions under degree 2, at a slightly higher cost and computation time.  

Future work will focus on advancing computational tractability of the proposed model for large-scale real-world power systems through decomposition methods while preserving the nonlinear physics of the network models.

% \section*{Acknowledgments}
% This should be a simple paragraph before the References to thank those individuals and institutions who have supported your work on this article.

% {\appendix[Proof of the Zonklar Equations]
% Use $\backslash${\tt{appendix}} if you have a single appendix:
% Do not use $\backslash${\tt{section}} anymore after $\backslash${\tt{appendix}}, only $\backslash${\tt{section*}}.
% If you have multiple appendixes use $\backslash${\tt{appendices}} then use $\backslash${\tt{section}} to start each appendix.
% You must declare a $\backslash${\tt{section}} before using any $\backslash${\tt{subsection}} or using $\backslash${\tt{label}} ($\backslash${\tt{appendices}} by itself
%  starts a section numbered zero.)}

%{\appendices
%\section*{Proof of the First Zonklar Equation}
%Appendix one text goes here.
% You can choose not to have a title for an appendix if you want by leaving the argument blank
%\section*{Proof of the Second Zonklar Equation}
%Appendix two text goes here.}

% \section{References}
% You can use a bibliography generated by BibTeX as a .bbl file.
%  BibTeX documentation can be easily obtained at:
%  http://mirror.ctan.org/biblio/bibtex/contrib/doc/
%  The IEEEtran BibTeX style support page is:
%  http://www.michaelshell.org/tex/ieeetran/bibtex/
 
%  % argument is your BibTeX string definitions and bibliography database(s)

\renewcommand{\bibfont}{\footnotesize}
\bibliographystyle{IEEEtran}
\bibliography{reference}

% Generated by IEEEtran.bst, version: 1.14 (2015/08/26)
\begin{thebibliography}{10}
\providecommand{\url}[1]{#1}
\csname url@samestyle\endcsname
\providecommand{\newblock}{\relax}
\providecommand{\bibinfo}[2]{#2}
\providecommand{\BIBentrySTDinterwordspacing}{\spaceskip=0pt\relax}
\providecommand{\BIBentryALTinterwordstretchfactor}{4}
\providecommand{\BIBentryALTinterwordspacing}{\spaceskip=\fontdimen2\font plus
\BIBentryALTinterwordstretchfactor\fontdimen3\font minus \fontdimen4\font\relax}
\providecommand{\BIBforeignlanguage}[2]{{%
\expandafter\ifx\csname l@#1\endcsname\relax
\typeout{** WARNING: IEEEtran.bst: No hyphenation pattern has been}%
\typeout{** loaded for the language `#1'. Using the pattern for}%
\typeout{** the default language instead.}%
\else
\language=\csname l@#1\endcsname
\fi
#2}}
\providecommand{\BIBdecl}{\relax}
\BIBdecl

\bibitem{AC_scopf_def}
M.~I. Alizadeh and F.~Capitanescu, ``Affinely adjustable robust optimization for constraint filtering in ac security constrained optimal power flow under uncertainties,'' \emph{IEEE Trans. Power Syst.}, vol.~40, no.~1, pp. 1118--1129, 2024.

\bibitem{stochastic_scopf}
M.~Alizadeh and F.~Capitanescu, ``A tractable linearization-based approximated solution methodology to stochastic multi-period ac security-constrained optimal power flow,'' \emph{IEEE Trans. Power Syst.}, vol.~38, no.~6, pp. 5896--5908, 2022.

\bibitem{SCOPF_review}
F.~Capitanescu, J.~M. Ramos, P.~Panciatici, D.~Kirschen, A.~M. Marcolini, L.~Platbrood, and L.~Wehenkel, ``State-of-the-art, challenges, and future trends in security constrained optimal power flow,'' \emph{Electric power syst. res.}, vol.~81, no.~8, pp. 1731--1741, 2011.

\bibitem{SCOPF_eff}
R.~Weinhold and R.~Mieth, ``Fast security-constrained optimal power flow through low-impact and redundancy screening,'' \emph{IEEE Trans. Power Syst.}, vol.~35, no.~6, pp. 4574--4584, 2020.

\bibitem{DC}
L.~Roald, S.~Misra, T.~Krause, and G.~Andersson, ``Corrective control to handle forecast uncertainty: A chance constrained optimal power flow,'' \emph{IEEE Tran. Power Syst.}, vol.~32, no.~2, pp. 1626--1637, 2016.

\bibitem{SOC_relax}
C.~J. Coffrin, ``Solving multi-contingency ac power flow problems with convex relaxations,'' Los Alamos National Laboratory, NM, USA, Tech. Rep., 2020.

\bibitem{SOC_relax2}
S.~I. Bugosen, R.~B. Parker, and C.~Coffrin, ``Applications of lifted nonlinear cuts to convex relaxations of the ac power flow equations,'' \emph{IEEE Trans. Power Syst.}, 2024.

\bibitem{QC_relax}
C.~Coffrin, H.~L. Hijazi, and P.~Van~Hentenryck, ``The qc relaxation: A theoretical and computational study on optimal power flow,'' \emph{IEEE Trans. Power Syst.}, vol.~31, no.~4, pp. 3008--3018, 2016.

\bibitem{ACDC}
G.~{Mohy-ud-din}, R.~Heidari, H.~Ergun, and F.~Geth, ``Ac--dc security-constrained optimal power flow for the australian national electricity market,'' \emph{Elect. Power Syst. Res.}, vol. 234, p. 110784, 2024.

\bibitem{ML}
S.~Park and P.~Van~Hentenryck, ``Self-supervised learning for large-scale preventive security constrained dc optimal power flow,'' \emph{IEEE Trans. Power Syst.}, vol.~40, no.~3, pp. 2205--2216, 2025.

\bibitem{SCOPF_review3}
L.~Roald and G.~Andersson, ``Chance-constrained ac optimal power flow: Reformulations and efficient algorithms,'' \emph{IEEE Trans. Power Syst.}, vol.~33, no.~3, pp. 2906--2918, 2017.

\bibitem{prob_SCOPF}
Y.~Chen, R.~Moreno, G.~Strbac, and D.~Alvarado, ``Coordination strategies for securing ac/dc flexible transmission networks with renewables,'' \emph{IEEE Trans. Power Syst.}, vol.~33, no.~6, pp. 6309--6320, 2018.

\bibitem{chance_SCOPF1}
E.~Karangelos and L.~Wehenkel, ``An iterative ac-scopf approach managing the contingency and corrective control failure uncertainties with a probabilistic guarantee,'' \emph{IEEE Trans. Power Syst.}, vol.~34, no.~5, pp. 3780--3790, 2019.

\bibitem{CVaR_robust_SCOCP}
R.~A. Jabr, ``Distributionally robust cvar constraints for power flow optimization,'' \emph{IEEE Trans. Power Syst.}, vol.~35, no.~5, pp. 3764--3773, 2020.

\bibitem{PCE}
D.~Shen, H.~Wu, B.~Xia, and D.~Gan, ``Polynomial chaos expansion for parametric problems in engineering systems: A review,'' \emph{IEEE Syst. J.}, vol.~14, no.~3, pp. 4500--4514, 2020.

\bibitem{gPCE}
A.~Koirala, T.~Van~Acker, R.~D’hulst, and D.~Van~Hertem, ``Uncertainty quantification in low voltage distribution grids: Comparing monte carlo and general polynomial chaos approaches,'' \emph{Sust. Energy Grids Networks}, vol.~31, p. 100763, 2022.

\bibitem{non_guassian}
K.~Yurtseven, A.~Koirala, H.~Ergun, and D.~Van~Hertem, ``Stochastic optimal power flow for hybrid ac/dc grids considering continuous non-gaussian uncertainty,'' \emph{Int. J. Elect. Power Energy Syst.}, vol. 170, p. 110828, 2025.

\bibitem{gPCE_C_SCOPF}
A.~Koirala, T.~Van~Acker, M.~U. Hashmi, R.~D'hulst, and D.~Van~Hertem, ``Chance-constrained optimization based pv hosting capacity calculation using general polynomial chaos,'' \emph{IEEE Trans. Power Syst.}, vol.~39, no.~1, pp. 2284--2295, 2023.

\bibitem{risk_gpce}
K.~Yurtseven, H.~Ergun, and D.~Van~Hertem, ``Risk-based stochastic optimal power flow for ac/dc grids using polynomial chaos expansion,'' in \emph{IEEE PES Innov. Smart Grid Techn. Europe}, 2024, pp. 1--5.

\bibitem{smooth_approx}
G.~{Mohy-ud-din}, R.~Heidari, F.~Geth, H.~Ergun, and S.~M. Muslem~Uddin, ``Ac-dc power systems optimization with droop control smooth approximation,'' in \emph{IEEE Australasian Universities Power Eng. Conf.}, 2024, pp. 1--6.

\end{thebibliography}

% \newpage

% \section{Biography Section}
% If you have an EPS/PDF photo (graphicx package needed), extra braces are
%  needed around the contents of the optional argument to biography to prevent
%  the LaTeX parser from getting confused when it sees the complicated
%  $\backslash${\tt{includegraphics}} command within an optional argument. (You can create
%  your own custom macro containing the $\backslash${\tt{includegraphics}} command to make things
%  simpler here.)
 
% \vspace{11pt}

% \begin{IEEEbiography}[{\includegraphics[width=1in,height=1.25in,clip,keepaspectratio]{fig1}}]{Michael Shell}
% Use $\backslash${\tt{begin\{IEEEbiography\}}} and then for the 1st argument use $\backslash${\tt{includegraphics}} to declare and link the author photo.
% Use the author name as the 3rd argument followed by the biography text.
% \end{IEEEbiography}

% \vspace{11pt}

% \begin{IEEEbiographynophoto}{John Doe}
% Use $\backslash${\tt{begin\{IEEEbiographynophoto\}}} and the author name as the argument followed by the biography text.
% \end{IEEEbiographynophoto}

% \vfill

\end{document}